\documentclass{emulateapj}

\shorttitle{Extreme Feedback and Reionization}
\shortauthors{Heckman et al.}

\begin{document}


\title{Extreme Feedback and the Epoch of Reionization: \\
    Clues in the Local Universe}


\author{Timothy M. Heckman and Sanchayeeta Borthakur}
\affil{Center for Astrophysical Sciences, Department of Physics \& Astronomy, Johns Hopkins University,
    Baltimore, MD 21218}
\email{heckman@pha.jhu.edu}

\author{Roderik Overzier and Guinevere Kauffmann}
\affil{Max-Planck-Institute for Astrophysics, Garching, D-85741, Germany}

\author{Antara Basu-Zych}
\affil{Goddard Space Flight Center, Greenbelt, MD 20771}

\author{Claus Leitherer and Ken Sembach}
\affil{Space Telescope Science Institute, Baltimore, MD 21218}

\author{D. Chris Martin}
\affil{Astronomy Department, Caltech, Pasadena, CA 91125}

\author{R. Michael Rich}
\affil{Department of Physics \& Astronomy, UCLA, Los Angeles, CA 90095}

\author{David Schiminovich}
\affil{Department of Astronomy, Columbia University, New York, NY 10027}

\and 

\author{Mark Seibert}
\affil{Carnegie Institution of Washington Observatories, Pasadena, CA 91101}

\author{}

\author{}

\author{}

\author{}



\begin{abstract}

The source responsible for reionizing the universe at $z > 6$ remains uncertain. While an
energetically adequate population of star-forming galaxies may be in place, it is unknown
whether a large enough fraction of their ionizing radiation can escape into the
intergalactic medium. Attempts to measure this escape-fraction in intensely star-forming
galaxies at lower redshifts have largely yielded upper limits. In this paper we present new
HST COS and archival FUSE far-UV spectroscopy of a sample of eleven Lyman Break
Analogs (LBAs), a rare population of local galaxies that strongly resemble the high-z
Lyman Break galaxies. We combine these data with SDSS optical spectra and Spitzer
photometry. We also analyse archival FUSE observations of
fifteen typical UV-bright local starbursts. We find
evidence of small covering factors for optically-thick neutral gas in three cases. This is
based on two independent pieces of evidence: a significant residual intensity in the cores
of the strongest interstellar absorption-lines tracing neutral gas and a small ratio
of extinction-corrected H$\alpha$ to UV plus far-IR luminosities.
These objects represent three of the four LBAs that contain a young, very compact ($\sim 10^2$ pc),
and highly massive ($\sim10^9$ M$_{\odot}$)  dominant central object (DCO).  These
three objects also differ from the other galaxies in showing a significant amount of blueshifted
Ly$\alpha$ emission, which may be related to the low covering factor of neutral gas.
All four LBAs with DCOs
in our sample show extremely high velocity outflows of interstellar gas, with line
centroids blueshifted by about 700 km s$^{-1}$ and maximum outflow velocities reaching at
least 1500 km s$^{-1}$. We show that these properties are consistent with an outflow
driven by a powerful starburst that is exceptionally compact. We speculate that such
extreme feedback may be required to enable the escape of ionizing radiation from star
forming galaxies.

\end{abstract}





\section{Introduction}

Following the epoch of recombination at z $\sim$1000, the next major event in the
history of the universe was its reionization by an early generation of massive stars and/or
black holes. One of the major goals of modern cosmology is to understand this process:
what sources were responsible and when did it occur? Observations of the cosmic
microwave background with WMAP show that reionization occurred at a mean redshift
of 11.0$\pm$1.4 (Dunkley et al. 2009), while observations of absorption due to the
Lyman $\alpha$ forest in quasars show that reionization was not quite complete until a
redshift of about six (e.g. Fan et al. 2006).

Based on the observed luminosity functions of active galactic nuclei and star-forming
galaxies at the relevant redshifts, the latter appear to be the more energetically plausible
source for reionization (e.g. Bouwens et al. 2010). However, a major uncertainty in the
contribution of star forming galaxies is the poorly constrained fraction of ionizing
photons that are able to leak out of such galaxies into the intergalactic medium. Bouwens
et al. (2010) estimate that an escape fraction of 20\% (60\%) is required for consistency
with the WMAP results at the 2$\sigma$ (1 $\sigma$) level.

Many attempts have been made to measure this escape-fraction in star-forming galaxies
at low-redshift (Leitherer et al. 1995; Hurwitz et al. 1997; Deharveng et al. 2001;
Heckman et al. 2001 - hereafter H01; Bergvall et al. 2006; Grimes et al. 2009 -  hereafter G09), intermediate
redshift (Malkan et al. 2003, Siana et al. 2007,2010; Cowie et al. 2009) and high-
redshift (Steidel et al. 2001; Giallongo et al. 2002; Fernandez-Soto et al. 2003; Inoue et
al. 2005; Vanzella et al. 2010). For the most part, these measurements have only yielded
upper limits, although significant escaping Lyman continuum emission has been detected
in 17 out of a sample of 198 Lyman Break and Lyman $\alpha$ galaxies at $z >3$ (Iwata
et al. 2009), consistent with earlier results for a smaller sample of Lyman Break galaxies
(Shapley et al. 2006). In contrast, Vanzella et al. (2010)  found evidence for escaping Lyman
continuum in only 1 of 100 Lyman Break galaxies at z $\sim$ 4.

Why do only a minority of the high-z galaxies and none of the low-z galaxies show a significant
leakage of ionizing radiation?  In this regard, it is instructive to note that an HI column
density of only 1.6 $\times 10^{17}$ cm$^{-2}$ is required to produce unit optical depth at
the Lyman edge. The area-averaged column densities of cold gas in star-forming galaxies are
much larger, ranging from  $10^{21}$ cm$^{-2}$ in ordinary low-redshift galactic disks
to as high as  $10^{24}$ cm$^{-2}$ in local starbursts and high-z star-forming galaxies
(Kennicutt 1998; Genzel et al. 2010). The escape of ionizing radiation must then be
determined by the topology of the interstellar medium, which will in turn be strongly
influenced by feedback from massive stars. It may be that rather exceptional
circumstances are required to produce the required channels of very low column-density
neutral gas. Such processes have been extensively studied from a theoretical
perspective ovre the past decade (e.g. Dove et al. 2000; Clarke \& Oey 2002; Fujita et al. 2003;
Gnedin \& Kravtsov 2008; Wise \& Cen 2009; Razoumov \& Somer-Larson 2010; Yajima et al. 2010).

These considerations provide the motivation of the present paper: to find local examples
of galaxies in which a potentially significant fraction of the ionizing radiation may be
escaping, and to then exploit the relative ease of observing such bright local objects to
gain insight into the physical processes that facilitate this escape. To be able to assess the
cosmological implications of the results of such an approach, it is important to observe
local galaxies that most closely resemble the type of UV-bright high-redshift galaxies that
are candidates for the reionization of the universe.

We began such an investigation in Overzier et al. (2009  - hereafter O09), where we
reported on Spitzer photometry, HST imaging, and Sloan Digital Sky Survey (SDSS)
optical spectroscopy of a sample of 31 Lyman Break Analogs (LBAs). These LBAs are a
very rare population of local galaxies that strongly resemble the high-z Lyman Break
galaxies (Heckman et al. 2005; Hoopes et al. 2007; Overzier at al. 2008;2010; Basu-
Zych et al. 2007;2009). They are defined to have a range in far-UV luminosity
($> 2 \times 10^{10}$ L$_{\odot}$) and far-UV effective surface brightness
($> 10^9$ L$_{\odot}$ kpc$^{-2}$) that are similar to typical Lyman Break galaxies.   

We showed in O09 that the morphologies of LBAs as revealed in multi-band HST
images fall into two distinct categories. The first class (25 objects) consists of 
galaxies in which multiple bright
complexes of clumps and knots reside in a galaxy whose disturbed morphology
is suggestive of a major merger or strong tidal interaction.  The second class (six objects) 
consists of galaxies whose UV image 
is dominated by a single extremely massive ($M_* \sim$ one-to-several billion $M_{\odot}$)
and compact (radius of-order 100 pc) central object (a dominant central object, or DCO).
We pointed out
that the star-formation rates derived for these six galaxies based on the extinction-
corrected H$\alpha$ luminosities were systematically smaller than those derived from the
Spitzer mid-IR luminosity (by factors of $\sim$ 2 to 5), and speculated that a possible
explanation was that these galaxies were leaking ionizing radiation into the intergalactic
medium. 

In this paper we test this idea using new Hubble Space Telescope (HST) far-UV observations
with the Cosmic Origins Spectrograph  (COS) of a sample of eight LBAs from O09. We
will combine these new data with archival Far Ultraviolet Spectroscopic Explorer
(FUSE) data for three other LBAs and for fifteen more typical UV-bright local starbursts.
We will describe the sample selection in section 2 and the observations and data analysis
in section 3. We will present our results and consider their interpretation in section 4, and
summarize these results and their implications in section 5.

\begin{figure*}
\includegraphics[scale=8.5,angle=-0]{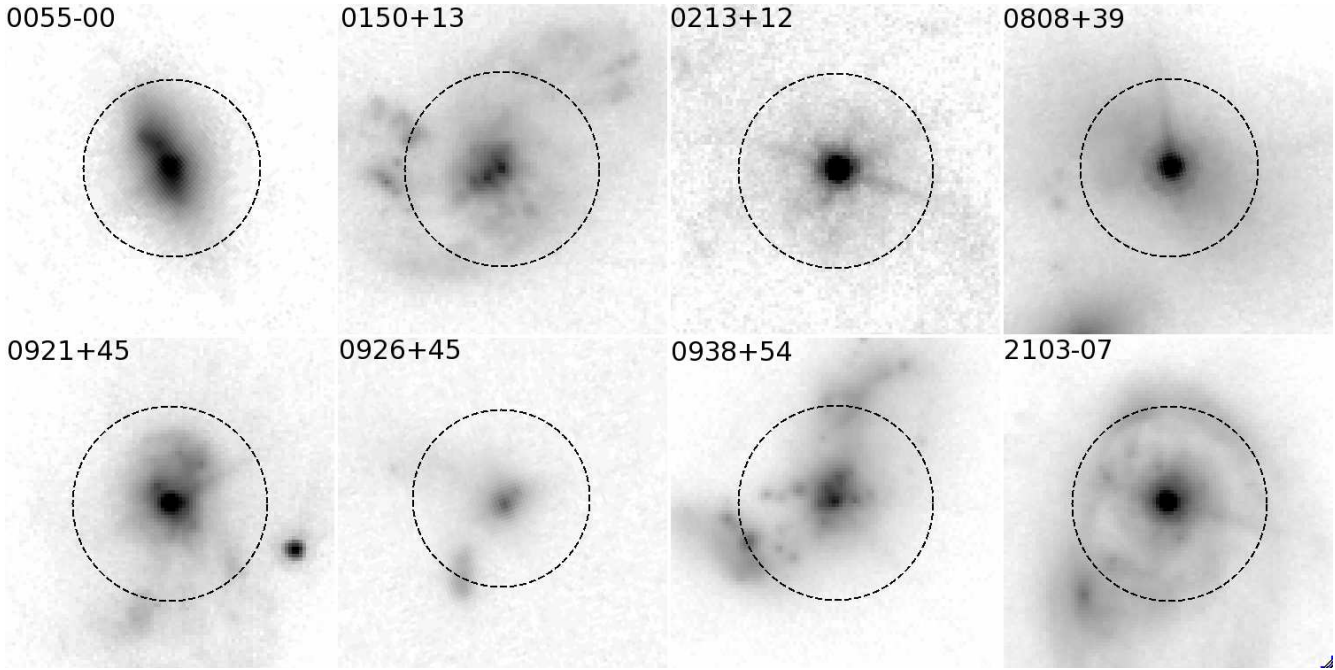} 
\caption{HST rest-frame optical images of the eight Lyman Break Anologs observed with COS. The circle indicates the 2.5 arcsec aperture of COS. For details on the images see Overzier et al. (2009). Note the diffraction spikes visible in the images of 0213+12, 0808+39, 0921+45, and 2103-07 produced by the barely-resolved Dominant Central Object.}
\end{figure*}

\section{Sample Selection}

We have previously used the Hopkins Ultraviolet Telescope and FUSE to search for the
direct evidence of the escape of ionizing radiation from local starbursts (Leitherer et al.
1995; H01; G09), but the modest sensitivity of these facilities was sufficient only to
observe three LBAs (none of which contained a DCO). With the recent installation of the
Cosmic Origins Spectrograph (COS) on the Hubble Space Telescope (HST), we are now
in a position to obtain high-quality far-UV spectra of a significantly larger sample of
these galaxies, including some with DCOs.

For the redshifts of the LBAs in our HST program ($z \sim$ 0.1 to 0.25), the sensitivity of COS is poor in the
rest-frame Lyman continuum. The sensitivity is much better at longer wavelengths, so in
this paper we follow H01 and G09 and use the residual intensity in the core of the
strongest far-UV interstellar absorption-lines from the neutral phase of the interstellar
medium as a probe of the intensity in the region below the Lyman edge. We will describe
this in more detail in section 3 below.

This approach requires the observations to be obtained through an aperture that encloses the
bulk of the starburst's emission (so that a global constraint is obtained). It also requires that the
width of instrumental spectral line-spread function is significantly smaller than the
characteristic velocity dispersion in the starburst (so that the interstellar absorption lines
are well resolved). Finally, the data need to have a high signal-to-noise ratio in the far-
UV continuum so that useful constraints are derived.

Based on these considerations, we will consider two data sets in this paper. The first
consists of the sample of eighteen galaxies studied by H01 and G09. These are local
starburst or star-forming galaxies observed with the Far-Ultraviolet Spectroscopic
Explorer (FUSE - Moos et al. 2000), each having signal-to-noise better than 4.6 per
0.078 \AA\  ($\sim$ 25 km s$^{-1}$) spectral element in the FUSE LiF1A channel (see
G09). The observations and properties of this sample are described in Table 1. Three of
these galaxies are in fact LBAs without a DCO:  Mrk 54 (Deharveng et al. 2001), Haro 11 (Grimes et al.
2007), and VV 114 (Grimes et al. 2006). However, most of these galaxies have considerably
lower UV luminosities and star-formation rates than the LBAs. They also
span a much broader range in galaxy mass and metallicity (see G09 and O09, and compare
Table 1 and 2).

The second is a sample of eight LBAs observed with the Cosmic Origins Spectrograph
(Froning \& Green 2009) on the Hubble Space Telescope (Program 11727: PI  T.
Heckman). These are members of a sample of 31 LBAs with HST UV images discussed
by O09 and were selected for spectroscopy based on a high UV flux through the COS
aperture and a compact UV size (so that the COS line-spread function is not significantly
degraded). These observations and the properties of this sample are listed in Table 2 and
HST images are shown in Figure 1.
In the discussion to follow we will refer to the eight LBAs with HST COS data and the
three LBAs with FUSE data as the LBA sample.  We will refer to the other fifteen
galaxies with FUSE data as the local starburst sample.

\begin{figure*}
\includegraphics[trim = 10mm 80mm 0mm 0mm, clip,scale=.95,angle=-0]{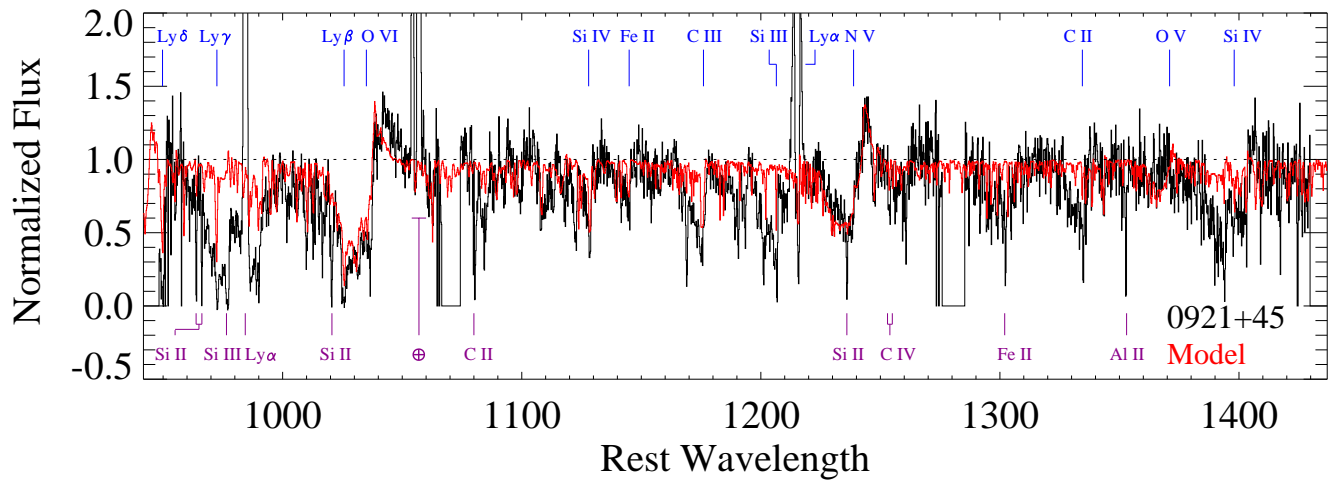} \\
\caption{The full COS G130M and G160M spectrum of LBA0921+45 is shown in black, while an illustrative Starburst99 model spectrum is shown in red (constant star formation for ten Myr with 0.4 solar metallicity and a Chabrier IMF). Both spectra have been normalized to unit flux (see text). Prominent spectral features in the starburst are labeled at the top, while Milky Way and geocoronal features are labeled at the bottom.}
\end{figure*}
\section{Observations and Data Analysis}

\subsection{The Data}

The observations and data reduction of the FUSE sample have been described in detail in
H01 and G09. We refer the reader to these papers. All the data were obtained through the
30 x 30 arcsec LWRS aperture, except for the cases of NGC 5253 and NGC 7714 (which
used the 4 x 20 arcsec  MDRS aperture). As shown by G09 these apertures encompass
most or all of the starburst in the far-UV. These spectra cover the observed wavelength
range from 905 to 1187  \AA\ (see Table 1 for the corresponding range in the rest-frame).
Depending on the angular size of the starburst in the far-UV, the
instrumental spectral resolution is $R  \sim$ 5000 to 14,000, corresponding to a velocity
dispersion of $\sigma \sim$ 9 to 25 km s$^{-1}$ (G09). In all cases the interstellar
absorption lines in the starbursts are well resolved.

For the HST-COS sample we have used the COS G130M and G160M gratings to obtain
spectra of our eight targets. As can be seen in Figure 1, the COS
aperture encompasses most or all of the galaxy.  We have retrieved these data from the HST MAST archive
after they have been processed through the standard COS pipeline. The merged spectra
cover a range from about 1160 to 1780 \AA\, with the corresponding range in the rest-
frame wavelength given for each sample member in Table 2. As with the FUSE data, the
instrumental spectral resolution depends on the angular size of the far-UV continuum
within the COS aperture. Our HST UV images (O09) yield sizes of 0.1 to 0.5 arcsec (FWHM),
and we estimate that the resulting spectral resolution is $R \sim$ 3000 to 13,000,
corresponding to a velocity dispersion of $\sigma \sim$ 10 to 43 km s$^{-1}$
(similar to the FUSE data).

For the new COS spectra we have fit each merged G130M and G160M spectrum with a Starburst99 (Leitherer et al. 1999) model spectrum
for a young stellar population with a metallcity similar to that of the corresponding LBA (Table 2). We did
not attempt to obtain a rigorous best-fit, but simply adopted a standard Chabrier (2003) IMF and assumed a constant
rate of star-formation. We varied the duration of the star-formation until we obtained (by eye)
a satisfactory fit to the strong stellar wind features. 
We then took the ratio of the model and the data
and fit the result with a low-order polynomial (avoiding spectral regions with strong emission or
absorption features). We then divided the data by the polynomial to obtain a normalized spectrum. An example
of this is shown in Figure 2.

\subsection{The Analysis Approach}

The opacity of the interstellar medium to Lyman continuum photons is due to two
primary sources: the photoelectric opacity of the neutral hydrogen and the opacity of
dust. The latter significantly affects the entire UV continuum in typical local starbursts
and high-z galaxies. We will focus first on measuring constraints on the
photoelectric opacity. Later we will comment on the importance of dust.

The only direct probe of the photoelectric opacity is the direct measurement of the
relative intensity in the continuum shortward of the Lyman edge. The FUSE data provide
clean access to this spectral region and adequate signal-to-noise in only 5 targets. G09
have used these data to derive upper limits to the escape of ionizing radiation for these
galaxies (see also Deharveng et al. 2001; Bergvall et al. 2006). The COS data do not
sample this spectral region in any galaxy. Thus, to gain insight into the escape of ionizing
radiation for our two samples we will follow H01 and G09 and use as probes the residual
intensity in the core of the strongest interstellar absorption lines that primarily arise in the
neutral interstellar medium.

For the FUSE data the strongest such line is CII$\lambda$1036.3 and for the HST-COS
data the strongest such line is CII$\lambda$1334.5. Thus in both samples we will use the
same ionic species as a tracer of the neutral ISM. \footnote{In principle, a significant fraction of the gas-
phase carbon in the neutral medium could be in the form of CI (which has an ionization
potential of 11.52 eV).  The CI$\lambda$1277.2 line should be comparable in strength to
the two CII lines in this case, however it is extremely weak in our COS spectra and in the
spectra of other local starburst galaxies (e.g. Heckman et al. 1998; Leitherer et al. 2002;
Schwartz et al. 2006; Leitherer et al. 2010) and high-z Lyman Break
galaxies (Shapley et al. 2003; Pettini et al. 2002; Steidel et al. 2010; Quider et al.
2009;2010; Dessauges-Zavadsky et al. 2010).} We have measured the relative residual
intensity in the core of the CII absorption line, following the same methodology as in
H01 and G09. The results are listed in Tables 1 (FUSE) and 2 (COS).

\begin{deluxetable*}{lccccccccc}
\tabletypesize{\scriptsize}
\tablecaption{FUSE Sample}
\tablewidth{0pt}
\tablehead{
\colhead{Name} & \colhead{LBA} & \colhead{v$_{sys}$} & \colhead{Range} & \colhead{F$_{esc,rel}$} &
\colhead{F$_{esc,abs}$} & \colhead{SFR} & \colhead{log M$_*$} &
\colhead{$\Delta$$v_{cen}$} & \colhead{$\Delta$$v_{max}$} 
}

\startdata

HARO 11 & Y & 6175 & 887 to 1163 & $<$ 0.08 & $<$ 0.006 & 26 & 10.2 & -180 & -450 \\
VV114 & Y & 6016 & 887 to 1164 & $<$ 0.02 & $<$0.001 & 66 & 10.8 & -150 & -650 \\
NGC 1140 & N & 1501 & 900 to 1181 & $<$ 0.04 & $<$ 0.014 & 0.83 & 9.4 & -30 & -400 \\
SBS 0335-052 & N & 4043 & 893 to 1171 & $<$ 0.18 & $<$ 0.18 & 0.32 & 7.8 & 30 & -130 \\
TOL 0440-381 & N & 12291 & 869 to 1140 & $<$ 0.04 & $<$ 0.016 & 5.0 & 10.0 & -120 &
-450 \\
NGC 1705 & N & 632 & 903 to 1184 & $<$ 0.01 & $<$ 0.008 & 0.16 & 8.6 & -12 & --- \\
NGC 1741 & N & 4037 & 893 to 1171 & $<$ 0.01 & $<$ 0.002 & 6.0 & 9.7 & -35 & -500 \\
IZW 18 & N &  751 & 903 to 1184 & $<$ 0.1 & $<$ 0.1 & 0.016 & 7.1 & -3 & -150 \\
NGC 3310 & N & 983 & 902 to 1183 & $<$ 0.05 & $<$ 0.005 & 2.8 & 9.8 & -170 & -600 \\
HARO 3 & N & 963 & 902 to 1183 & $<$ 0.005 & $<$ 0.001 & 0.41 & 8.9 & 10 & -150 \\
NGC 3690 & N & 3121 & 896 to 1175 & $<$ 0.01 & $<$ 0.001 & 82 & 10.9 & -120 & -900 \\
NGC 4214 & N & 291 & 904 to 1186 & $<$ 0.01 & $<$ 0.002 & 0.13 & 9.0 & -40 & -250 \\
MRK 54 & Y & 13450 & 866 to 1136 & $<$ 0.09 & $<$ 0.05 & 21 & 10.4 & -90 &  -550 \\ 
M 83 & N & 517 & 903 to 1185 & $<$ 0.06 & $<$ 0.01 & 3.5 & 10.7 & 9 & --- \\
NGC 5253 & N & 405 & 904 to 1185 & $<$ 0.06 & $<$ 0.02 & 0.33 & 9.3 & 17 & --- \\
IRAS 19245-4140 & N & 2832 & 897 to 1176 & $<$ 0.21 & $<$0.11 & 2.1 & 9.1 & -60 & -350 \\
NGC 7673 & N & 3392 & 895 to 1174 & $<$ 0.1 & $<$ 0.017 & 4.8 & 10.0 & -80 & -350 \\
NGC 7714 & N & 2803 & 897 to 1176 & $<$ 0.03 & $<$ 0.01 & 6.9 &  10.2 & -60 & -500 \\

\enddata


\tablecomments{Column 1: Galaxy name. Column 2: Is the galaxy a Lyman Break Analog? See O09 for a definition. 
Column 3: Galaxy systemic velocity (km s$^{-1}$) from G09 or H01. Column 4: Wavelength range covered by the FUSE data
in the galaxy rest-frame (\AA). Column 5: The relative escape fraction of ionizing radiation based on the residual relative
intensity in the CII$\lambda$1036.3 line - see text for details. Column 6: The absolute escape fraction including the effect
of attenuation due to dust opacity - see text for details. Column 7: The star formation rate ($M_{\odot}$ year$^{-1}$), based
on the sum of the total infrared and
far-ultraviolet luminosities and assuming a Chabrier 2003 IMF - see G09 for details. Column 8: The log of the galaxy stellar mass
($M_{\odot}$) based on the 2MASS K-band luminosity and assuming a K-band mass-to-light ratio of 0.4 (Bell \& de Jong 2001). Column 9:
The centroid velocity of the strong interstellar absorption-lines with respect to the galaxy systemic velocity (km sec$^{-1}$). See G09
for details. Column 10: The maximum outflow velocity (in km sec$^{-1}$), following Steidel et al. (2010).} 

\end{deluxetable*}

\begin{deluxetable*}{lccccccccccccc}
\tabletypesize{\scriptsize}

\tablecaption{COS Sample}
\tablewidth{0pt}
\tablehead{
\colhead{Name} & \colhead{DCO} & \colhead{z$_{sys}$} & \colhead{Range} & \colhead{F$_{esc,rel}$} &
\colhead{F$_{esc,abs}$} & \colhead{SFR$_{IR+UV}$} & \colhead{SFR$_{H\alpha}$} & \colhead{log M$_*$} &
\colhead{$\Delta$$v_{cen}$} & \colhead{$\Delta$$v_{max}$} & \colhead{R$_{eqw}$} & \colhead{$\beta$} & \colhead{[O/H]}
}

\startdata

0055-00 & N & 0.16744 & 970 to 1513 & $<$0.05 & $<$0.01 & 29 & 23 & 9.7 & -230 & 
-640 & -0.21 & -1.5 & -0.41 \\

150+13 & N & 0.14668 & 1022 to 1572 & $<$0.1  & $<$0.01 & 56 & 19 & 10.3 & -150 &
-480 & -0.78 & -1.6 & -0.30 \\

0213+12 & Y & 0.21902 & 945 to 1440 &  0.35 & 0.05 &  55  & 6.7 &  10.5 & -770 &
-1500 & 0.36 & -0.9 & 0.05 \\

0808+39 & Y & 0.09123 & 1038 to 1607 & 0.3 & 0.12  & 11 & 3.7 & 9.8 & -680 &
-1500 & 0.27 & -1.0 & 0.05 \\

0921+45 & Y & 0.23499 & 948 to 1430 & 0.3 & 0.04 & 84 & 26 & 10.8 & -650 & -1500
& 0.75 & -1.4 & -0.02 \\

0926+45 & N & 0.18072 & 968 to  1515 & 0.6: & 0.4: & 10 & 13 & 9.1 & -300 & -700 &
0.14 & -2.3 & -0.60 \\

0938+54 & N &0.10208 & 1029 to 1614 & $<$0.1 & $<$0.05 & 13 & 13 & 9.4 & -90 &
-610 & -0.17 & -2.1 & -0.50 \\

2103-07 & Y & 0.13689 & 1030 to 1585 & $<$0.05 & $<$0.01  & 63 & 41 &  10.9 & -720 &
-1500 & 0.04 & -0.5 & 0.01 \\

\enddata


\tablecomments{Column 1: Galaxy name; Column 2: Does the galaxy contain a Dominant Central Object (DCO)? See O09 for details. Column 3: The galaxy
redshift from Overzier et al. (2011). Column 4: The wavelength range covered by the COS data
in the galaxy rest-frame (\AA). Column 5: The relative escape fraction of ionizing radiation based on the residual relative
intensity in the CII$\lambda$1334.5 line. In the case of 0926+45 the CII line has a significant residual intensity, but there is no deficiency of H$\alpha$
emission (unlike the cases of 0213+12, 0808+39, and 0921+45). Only these three latter objects are good candidates for the escape of ionizing radiation. 
See text for details. Column 6: The absolute escape fraction including the effect
of attenuation due to dust opacity - see text for details. Column 7: The star formation rate ($M_{\odot}$ year$^{-1}$), based
on the sum of the total infrared and
far-ultraviolet luminosities and assuming a Chabrier 2003 IMF - see Overzier et al. 2011 for details.
Column 8: The star formation rate ($M_{\odot}$ year$^{-1}$), based
on the extinction-corrected H$\alpha$ luminosity measured from the SDSS spectra - see Overzier et al. 2011 for details. Column 9: 
The log of the galaxy stellar mass
($M_{\odot}$). See O09 for details. Column 10: The centroid velocity of the SiIII$\lambda$1206.5 interstellar absorption-line
 with respect to the galaxy systemic velocity (km sec$^{-1}$).
Column 11: The maximum outflow velocity seen in the SiIII$\lambda$1206.5 line (in km sec$^{-1}$), following Steidel et al. (2010). Column 12:
The ratio of the equivalent widths of the blueshifted and redshifted components of the Ly$\alpha$ line. Values near one imply significant
blueshifted Ly$\alpha$ emission. See text for details. Column 13: The spectral slope of the UV continuum defined by $F_{\lambda} \propto
\lambda^{\beta}$. An unreddened starburst will have $\beta \sim$ -2.5 to -2. Six of the eight LBAs suffer significant dust reddening.
Column 14: The log of gas-phase O/H ratio relative to solar (Asplund et al. 2009). See Overzier et al. (2009) for details.}

\end{deluxetable*}

\section{Results \& Interpretation}

\begin{figure*}
\includegraphics[trim = 0mm 0mm 0mm 0mm, clip,scale=.7,angle=-0]{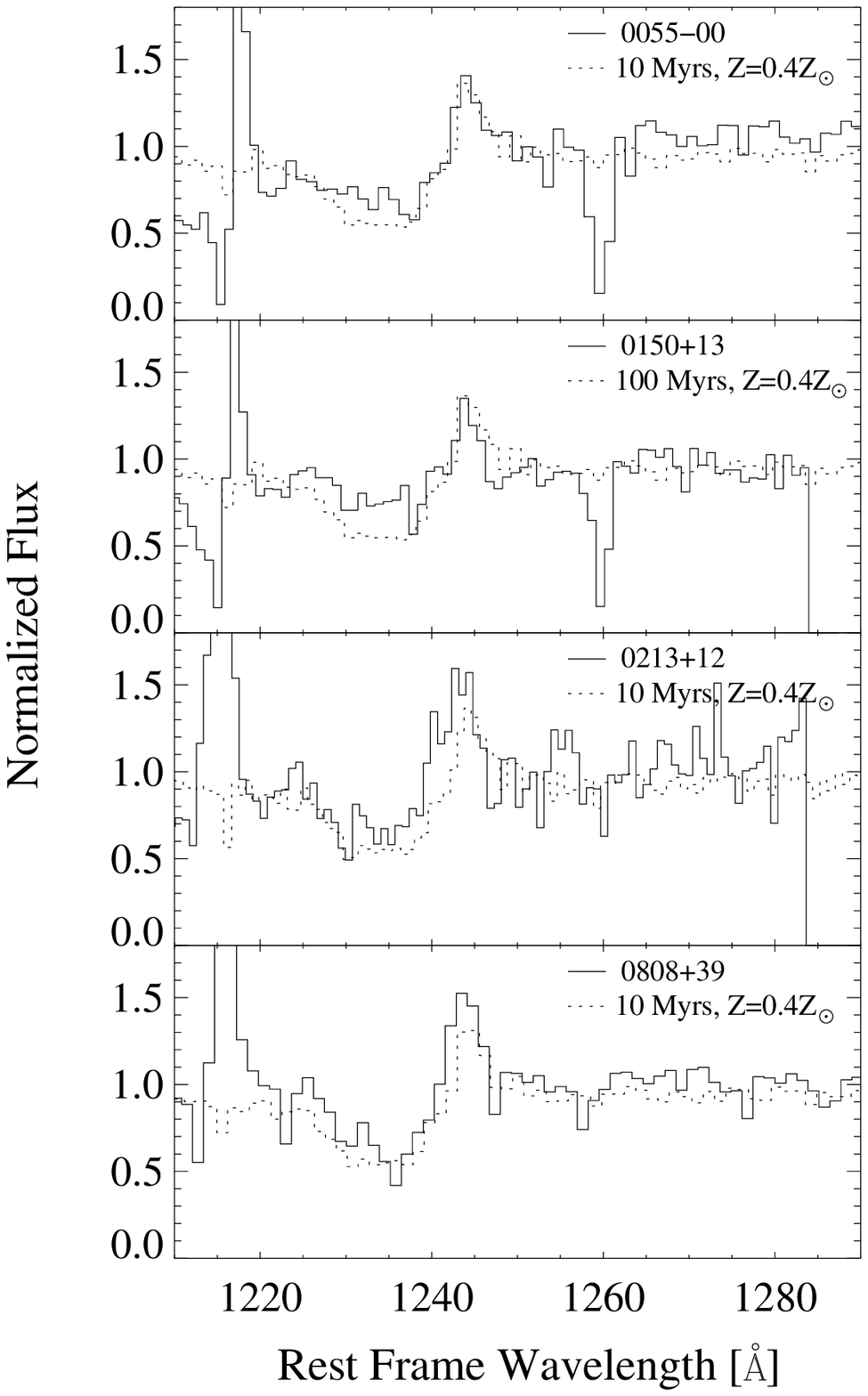} 
\includegraphics[trim = 0mm 0mm 0mm 0mm, clip,scale=.7,angle=-0]{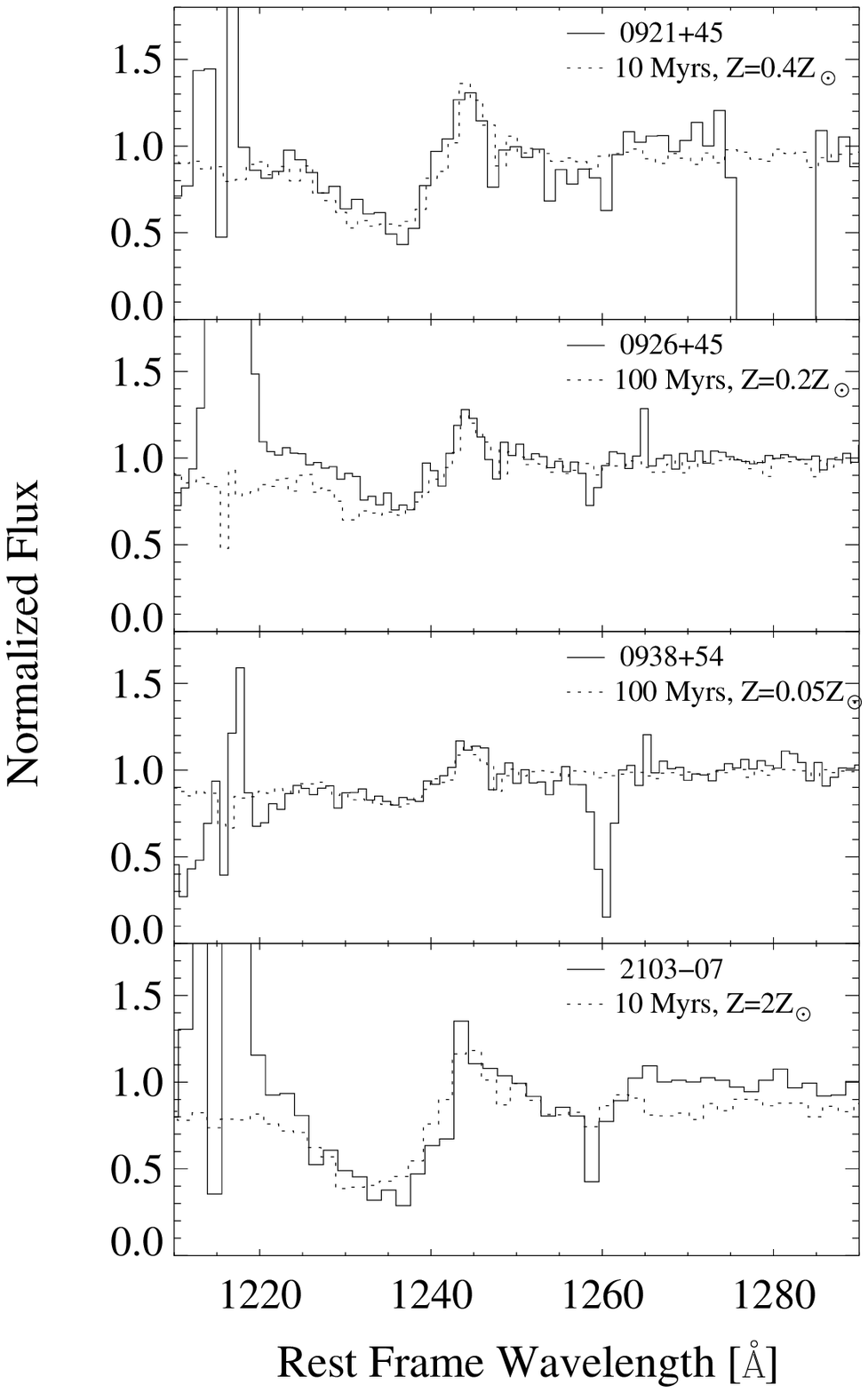} 
\caption{Spectra of the NV$\lambda$1240 stellar wind line in the eight Lyman Break Analogs. The dotted lines are illustrative models from Starburst99 that show that these spectra are consistent with the presence of a significant population of ionizing (O) stars. The models assume a rate of constant star-formation for either 10 or 100 Myr (as indicated), a Chabrier IMF, and a metallicity as indicated.}
\end{figure*}

\subsection{The Ionization Source}

\begin{figure*}
\includegraphics[trim = 0mm 0mm 0mm 0mm, clip,scale=.7,angle=-0]{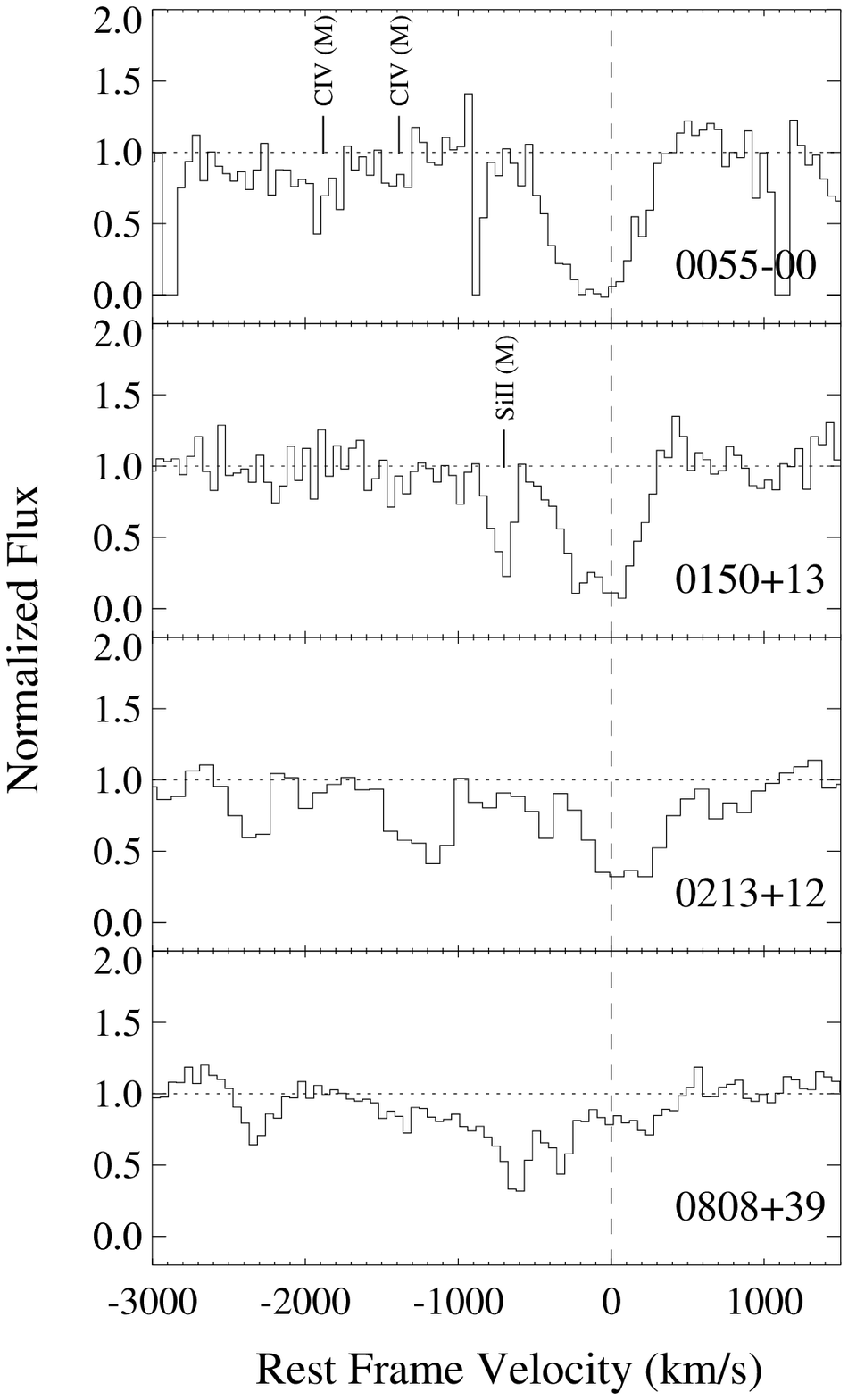} 
\includegraphics[trim = 0mm 0mm 0mm 0mm, clip,scale=.7,angle=-0]{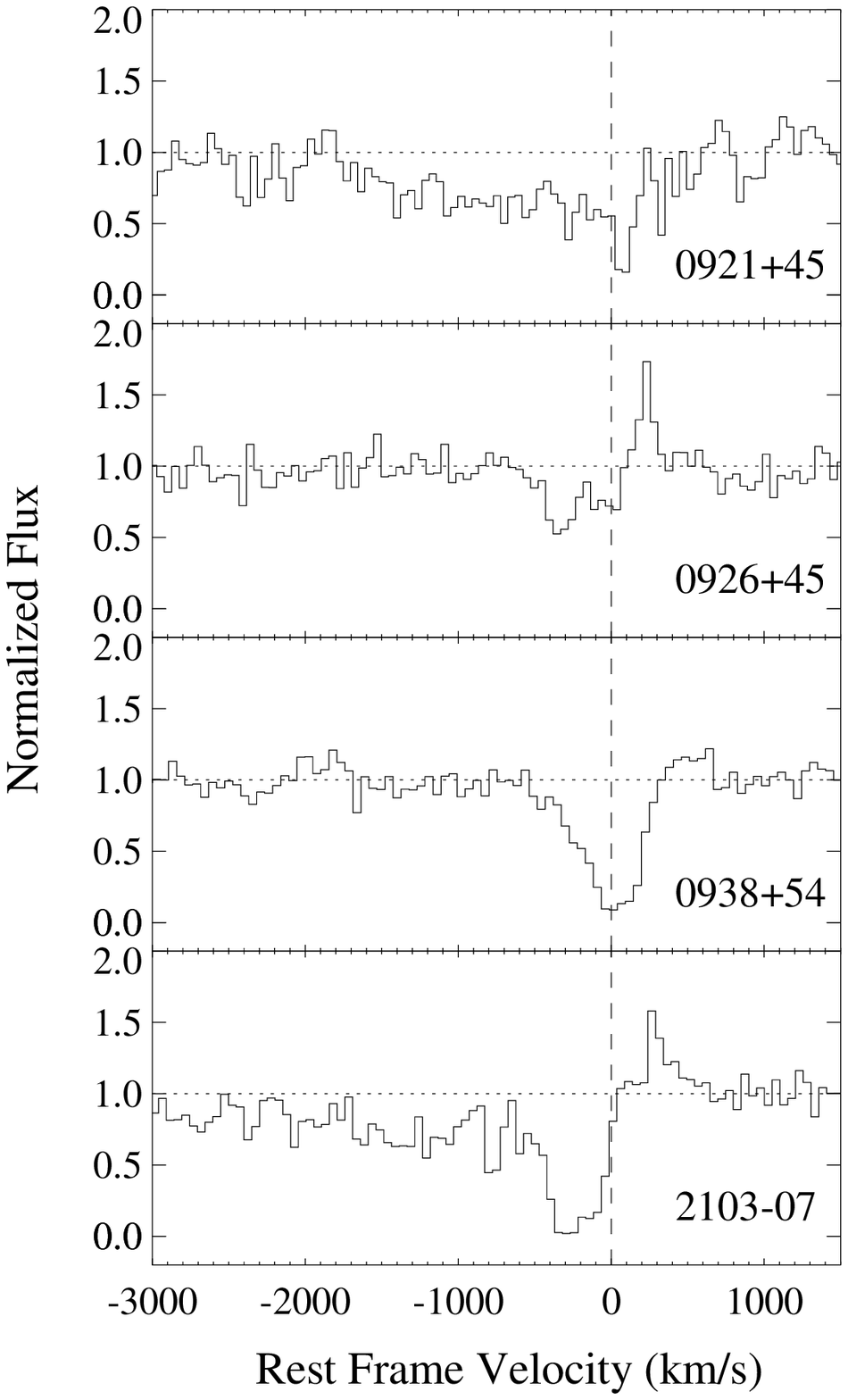} 
\caption{Absorption-line profiles of the CII$\lambda$1334.5 feature. The dotted line indicates the galaxy systemic velocity
derived from SDSS spectra. Contaminating foreground Milky Way features are noted. There is a significant relative residual intensity
in the line profile in four cases (0213+12, 0808+39, 0921+45, and 0926+45).}
\end{figure*}

Before considering constraints on the escape of ionizing radiation, it is important
to establish that an ionizing population of massive stars (O stars) is in fact present
in our targets. The strongest spectroscopic signature of these stars is provided by the
broad P-Cygni profiles in the resonance transitions of highly ionized species that
arise in the stellar winds (e.g. Robert et al. 1993; 2003). 

The strongest such feature that is present in all our COS spectra is NV$\lambda$1240.
In Figure 3 we show the NV profile for our eight targets and overplot the generic examples
of Starburst99 model spectra described above.
We defer to a future paper a detailed discussion of the properties of the stellar
populations in the LBAs. Here we merely point out that the far-UV spectra of the LBAs are
all consistent with a normal population of ionizing stars expected for a strong starburst.

\subsection{Constraints on the Escape of Ionizing Radiation}

In Figure 4 we show the CII$\lambda$1334.5 absorption-line profiles for our COS data
on eight LBAs. The corresponding profiles for the CII$\lambda$1036.3 line from the
FUSE data can be found in H01 and G09. A striking result is that there is a significant
residual intensity in the absorption-line profiles in four LBAs
(LBA0213+12, LBA0808+39, LBA0921+45, and LBA0926+45). The first three objects all contain
a DCO.
These results contrast strongly with the results in H01 and G09  in which
the CII$\lambda$1036.3 absorption-line in the FUSE data is black or nearly black at line
center in all cases.  

\begin{figure}
\includegraphics[trim = 10mm 0mm 0mm 0mm, clip,scale=.75,angle=-0]{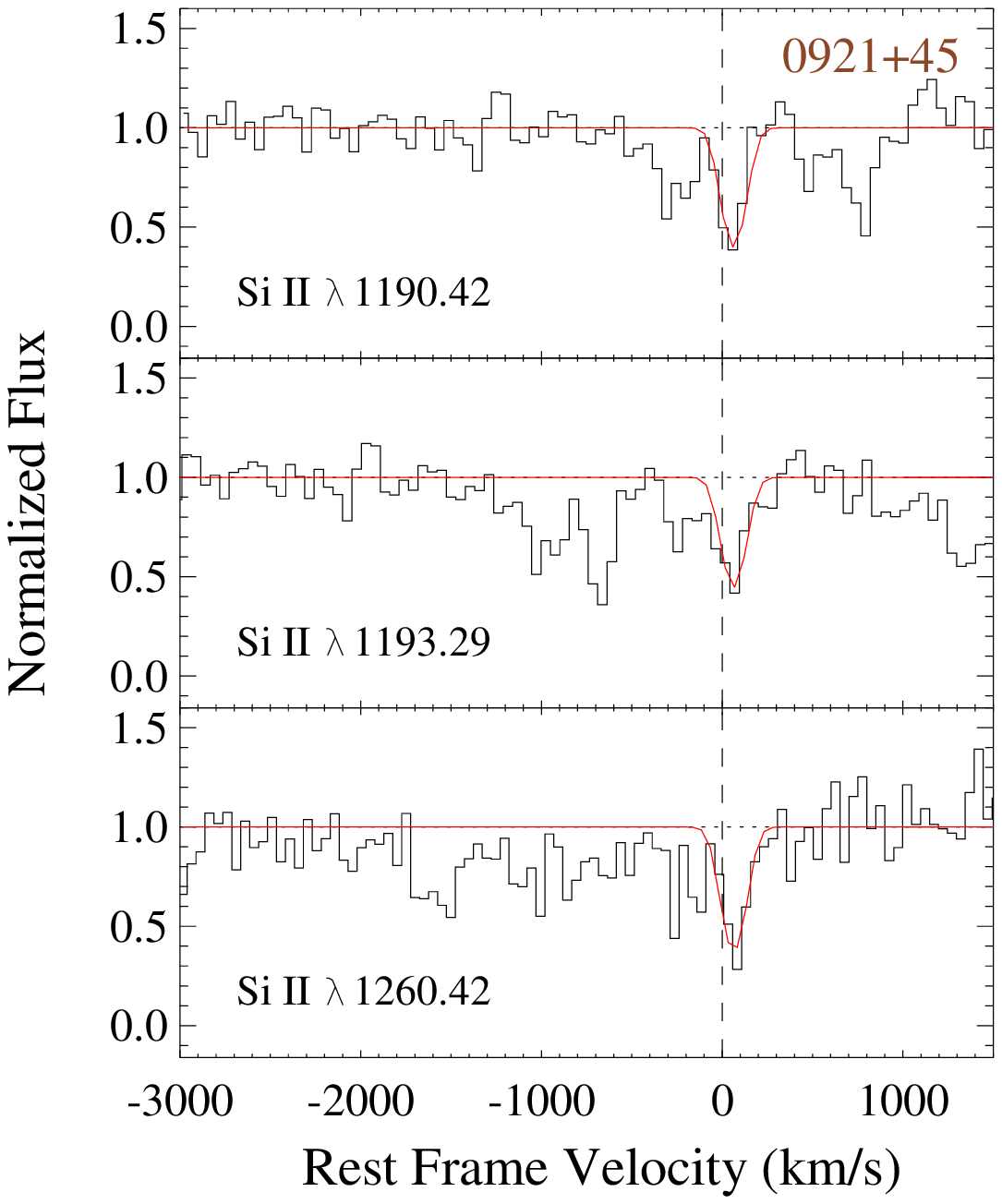} 

\caption{The SiII 1190.4, 1193.3, and 1260.4 absorption-line profiles for LBA0921+45 all plotted on the same scale. In each panel the red line shows a fit of a single gaussian to the strongest component of the profile. This was a joint simultaneous fit to all three features, which were forced to have a common centroid and line width, but a free line depth. The similar fitted depth of the feature in all three cases shows that the data imply gas that is optically thick, but does not fully cover the continuum source (a picket-fence model). See the text for details.}
\end{figure}

To understand our approach in using the CII line as a probe of the Lyman continuum, let
us consider two limiting idealized cases. In the first, we assume a picket-fence model in
which the far-UV source is surrounded by a population of clouds that are optically thick
in both the absorption line and the Lyman continuum and that cover a fraction $f_c$ of
the sky as seen by the source. In this case the residual relative intensity at the core of the
line ($I_0$) is just $I_0/I_{cont}  = 1 - f_c$. Thus, a significant residual intensity in the line core implies
a significant escape of Lyman continuum radiation.

 In the second case we assume that the far-UV source is surrounded by a uniform shell
with unit covering factor. As discussed in H01, in this case the optical depth in the ISM
just below the Lyman edge is related to the optical depth at the core of the absorption-line
by:\\

1)  $\tau_{Ly}/\tau_0 =  R Z_{gas}^{-1} (\sigma_{gas}/100$ km s$^{-1})$\\

where $Z_{gas}$ is the gas-phase carbon abundance scaled to the solar value of
$log(C/H) + 12 = 8.43$ (Asplund et al. 2009), we assume that CII is the dominant ionization
state in the neutral gas, $\sigma_{gas}$ is the velocity dispersion
in the neutral gas, and $R =$ 17 and 12 for the 1036.3 and 1334.5 lines
respectively (Morton 2003). Since the optical depth at the Lyman edge is over an order-
of-magnitude larger than in the line core, in this idealized case a significant residual
intensity in the core of the absorption lines is a necessary but not sufficient condition for
the escape of ionizing radiation.

\begin{figure*}
\includegraphics[trim = 0mm 0mm 0mm 0mm, clip,scale=.70,angle=-0]{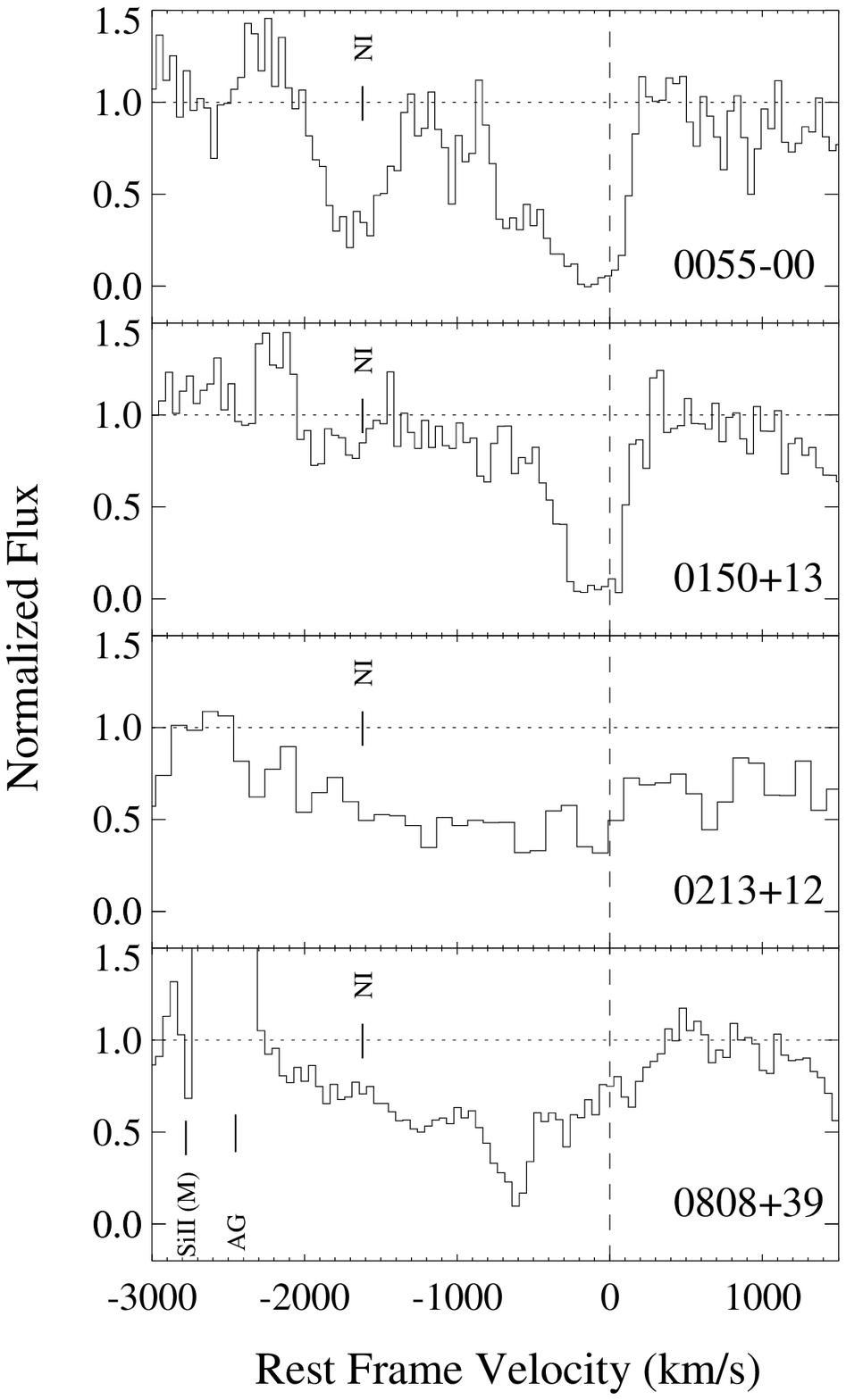} 
\includegraphics[trim = 0mm 0mm 0mm 0mm, clip,scale=.70,angle=-0]{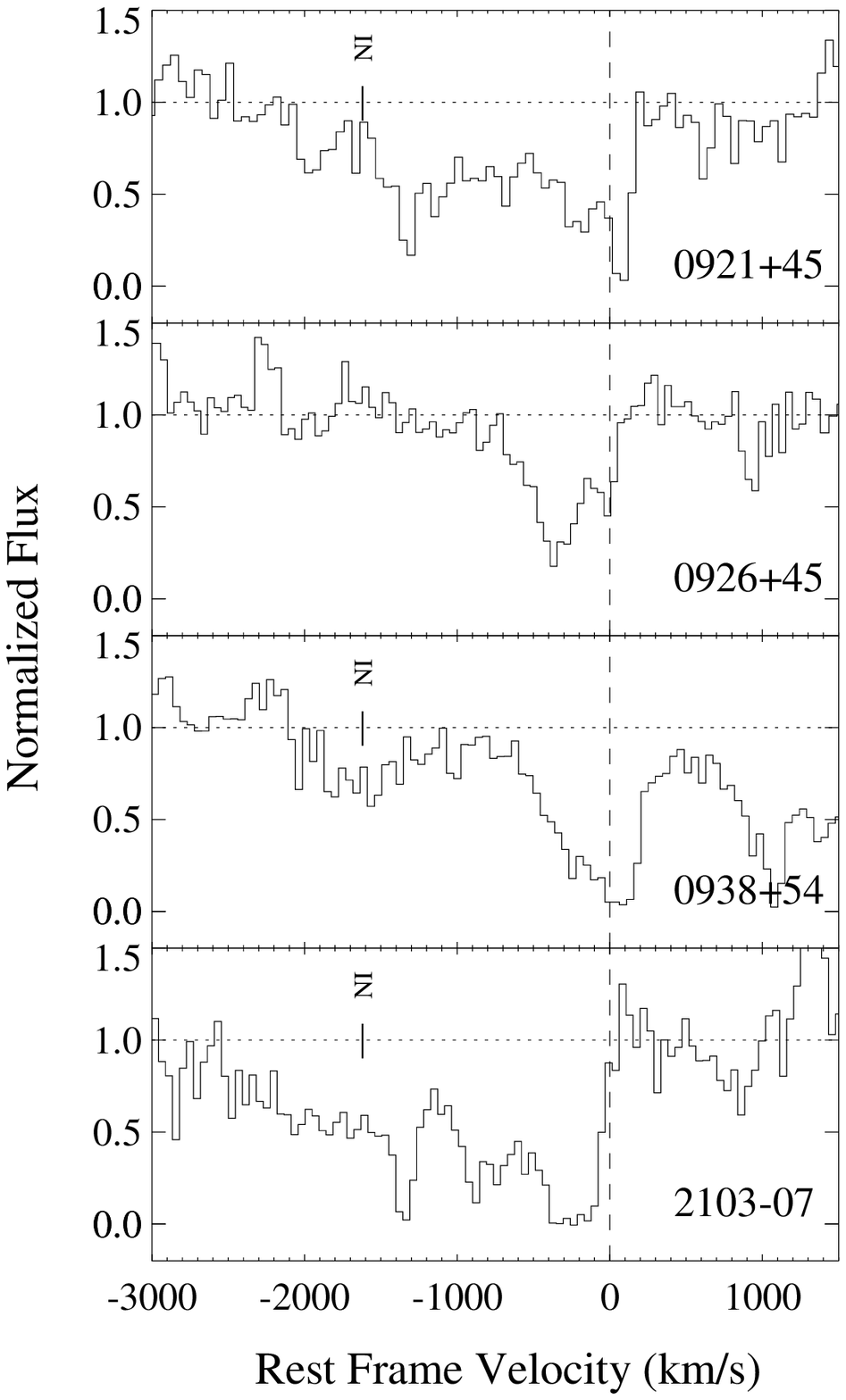} 
\caption{Absorption-line profiles of the SiIII$\lambda$1206.5 feature. Contaminating foreground Milky Way and air-glow features are noted. Highly blueshifted absorbing material is detected up to at least -1500 km s$^{-1}$ in the four cases with a Dominant Central Object (see text).  Beyond this velocity there is possible contamination by the NI$\lambda$1200.0 feature (as marked).}
\end{figure*}

To test these two models, we have used the set of SiII absorption-lines available in the
COS data. Like CII, SiII is the dominant ion of silicon in the neutral ISM.  By using lines
that all arise in the same ionic species, but which span a range in oscillator strengths, a
comparison of the absorption profiles among these lines allows us to estimate the optical
depth and covering factor of the gas (e.g. Heckman et al. 2000; Rupke et al. 2002; Quider
et al. 2009; Erb et al. 2010).
In the case of optically-thick clouds that do not fully cover the far-UV source, the line
profiles of the different transitions should look the same. In the case of a uniform shell,
the residual intensity will increase with decreasing oscillator strength $f$ (since
$\tau$$_0 \propto \lambda$$f$). We will use the following transitions (with the
respective values of $log \lambda$$f$ from Morton 2003): Si$\lambda$II1260.4 (-1.32),
1193.3 (-1.65), and 1190.4 (-1.95). \footnote{Unfortunately the SiII 1526 line (-2.18) is accessed
in only one object and the
1304.4 line (-2.44) is badly blended with the SiIII$\lambda$1301.2,1303.3 stellar
photospheric features (Walborn et al. 1995; Massa 1989) and the adjacent
OI$\lambda$1302.2 interstellar line).}  

For each galaxy we simultaneously fit a single gaussian profile to the three different SiII lines, forcing the
three lines to share a common centroid and width but allowing the normalization
(depth) of each line to vary freely.
In all four galaxies the results strongly favor the picket fence model with an implied
covering factor of optically-thick clouds of about 40\% for LBA0213+12 and LBA0926+45,
50\% for LBA0808+39, and 60\% for LBA0921+45. We show these fits for the case of LBA0921+45
in Figure 5.\footnote{For the other four LBAs the data imply
that the SiII lines are optically thick but fully cover the far-UV continuum source.}
In principle, the sightlines that avoid the
clouds could be optically-thin to the metal lines but still be optically-thick to the Lyman
continuum (equation 1). Thus, confirmation that there is significant escaping ionizing radiation
will require direct observations below the Lyman edge in these cases.

In the mean time we can consider whether there is other evidence for the escape of ionizing radiation
in these four LBAs. In
O09 we highlighted the relative weakness of the extinction-corrected H$\alpha$
emission-line in most of the LBAs with DCOs, and speculated that a significant fraction
of the ionizing radiation might be escaping. This information is listed in Table 2 where
we compare the star-formation rates derived from the extinction-corrected H$\alpha$
emission-line ($SFR_{H\alpha}$) to that obtained from the sum of the total-IR and far-
UV  luminosity ($SFR_{IRUV}$). It is noteworthy that three of the four objects above
in which we have inferred an incomplete covering of the far-UV source by neutral clouds
have the highest ratios of
$SFR_{IRUV}/SFR_{H\alpha}$: LBA0213+12 (a ratio of 8.2), LBA0808+39 (3.0), and
LBA0921+45 (3.2). All three of these contain a DCO. In contrast, the mean ratio 
of $SFR_{IRUV}/SFR_{H\alpha}$ for the other five galaxies is 1.4. {\it Thus, based
on two independent lines of evidence, we suggest that there is a significant
leakage of ionizing radiation from three of the eight LBAs in our sample.}

How do these results compare to what is seen at high redshift? There are four
gravitationally-lensed high-z galaxies that have been observed with spectral resolution
and signal-to-noise similar to our sample. \footnote{These strong interstellar metal absorption-lines from
neutral gas exhibit significant residual intensity in the high signal-to-noise stacked
spectra of high-z star-forming galaxies (e.g. Shapley et al. 2003; Steidel et al. 2010).
However, these data are obtained at relatively low spectral resolution ($R \sim 10^3$), so
that the line profiles are not well-resolved in individual objects (and could miss gas with a high covering factor
but low velocity dispersion). The stacking process itself could also smear out black, narrow
components found in the spectra of individual galaxies.}
Of these, the residual relative intensities of
CII$\lambda$1334.5, OI $\lambda$1302.2, and SiII $\lambda$ 1260.4 (the strongest
metal lines tracing the neutral ISM) are very small in MS 1512-cB58 (Pettini et al. 2002),
The Cosmic Eye (Quider et al. 2010), and the 8 O'Clock Arc (Dessauges-Zavadsky et
al. 2010), but large ($\sim$40\%) in The Cosmic Horseshoe (Quider et al. 2009). In fact,
the last object appears to a good example of the type of picket fence situation seen in the
DCOs in our sample.

As discussed above, dust can also be a significant source of opacity to ionizing radiation
in galaxies. Following Shapley et al. (2006) and G09, we can distinguish between the
absolute and relative escape fractions. The former is the actual fraction of ionizing
photons that escape the galaxy, including the effect of dust. The latter neglects the effect
of dust and is defined as the ratio of the fraction of escaping ionizing photons to escaping
non-ionizing far-UV photons (e.g. it measures only the effect of the photoelectric opacity
of the gas). Our measurements discussed above provide information about the
relative escape fraction. Given that the ratio of far-IR to far-UV fluxes for the galaxies
with DCOs are of-order ten, the implied absolute escape fractions will be proportionately
smaller. We list our estimates for both the relative and absolute escape fractions in Tables
1 (FUSE) and 2 (COS).

\subsection{Extreme Feedback}

As can be seen in Figure 4, the CII$\lambda$1334.5 absorption-line profiles in all the LBAs are blueshifted with respect
to the galaxy systemic velocity, implying a large-scale outflow of gas. Further evidence
for outflows is provided by the broad blue-asymmetric wings seen on the optical emission-line profiles (O09).

The absorption-line profiles in Figure 4 show evidence for exceptionally high outflow speeds in the DCOs.
To further investigate this, we turn to the Si III
$\lambda$1206.5 feature. This is the strongest metal line that is accessed in all four DCO
spectra, and arises in the ionized gas. As shown in Figure 6, outflowing gas is detected at
extraordinarily high velocities in all four galaxies with DCOs. The flux-weighted line
centroid is blueshifted by about 700 km s$^{-1}$ in these objects (Table 2).  It is difficult
to determine the maximum outflow velocity because of  blending with the
NI$\lambda$1200 triplet, which lies $\sim$1600 km s$^{-1}$ blueward of the SiIII
line.
Conservatively, the maximum outflow speed seen in the DCOs reaches at least
1500 km s$^{-1}$.

\begin{figure*}

\includegraphics[trim = 0mm 0mm 0mm 0mm, clip,scale=.45,angle=-0]{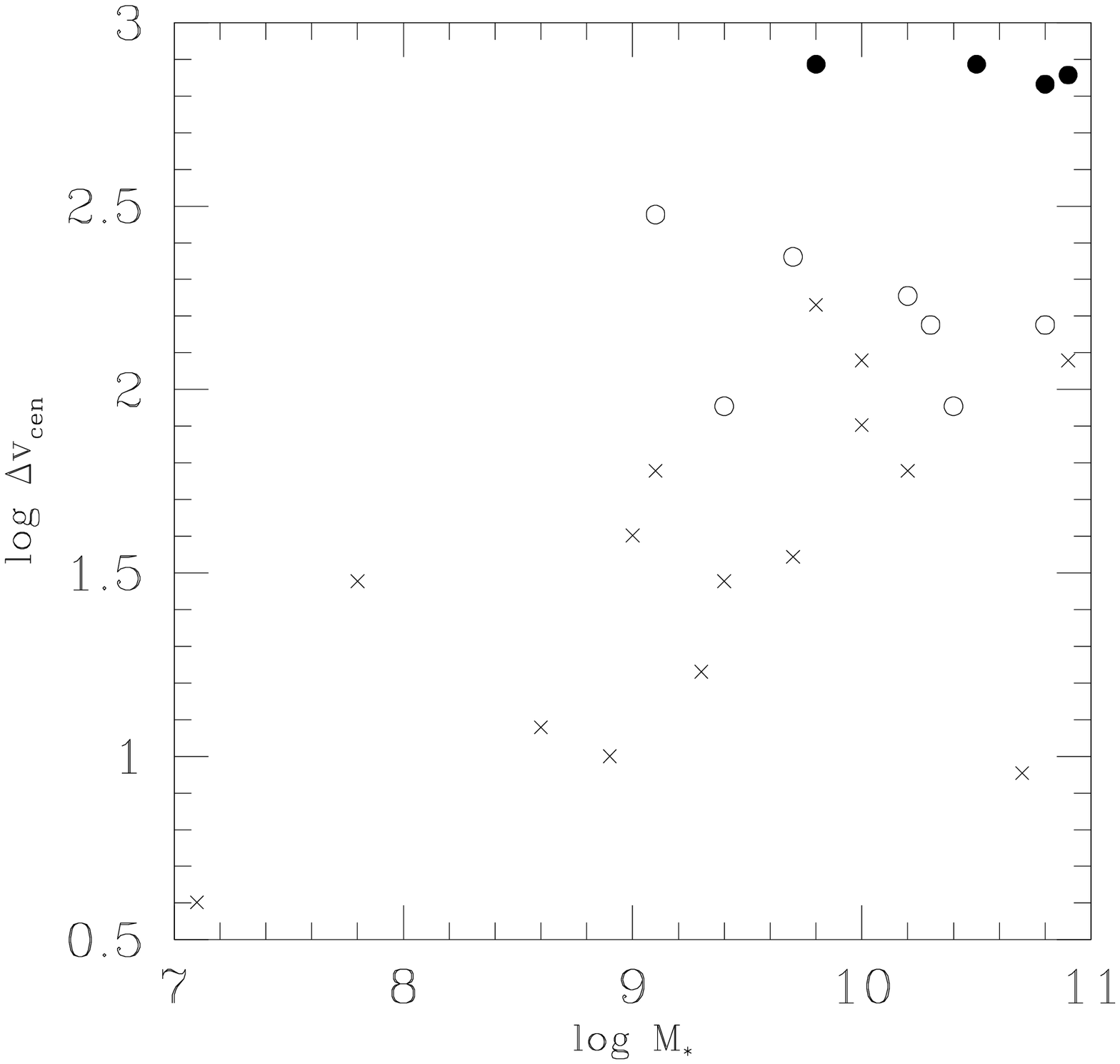} 
\includegraphics[trim = 0mm 0mm 0mm 0mm, clip,scale=.45,angle=-0]{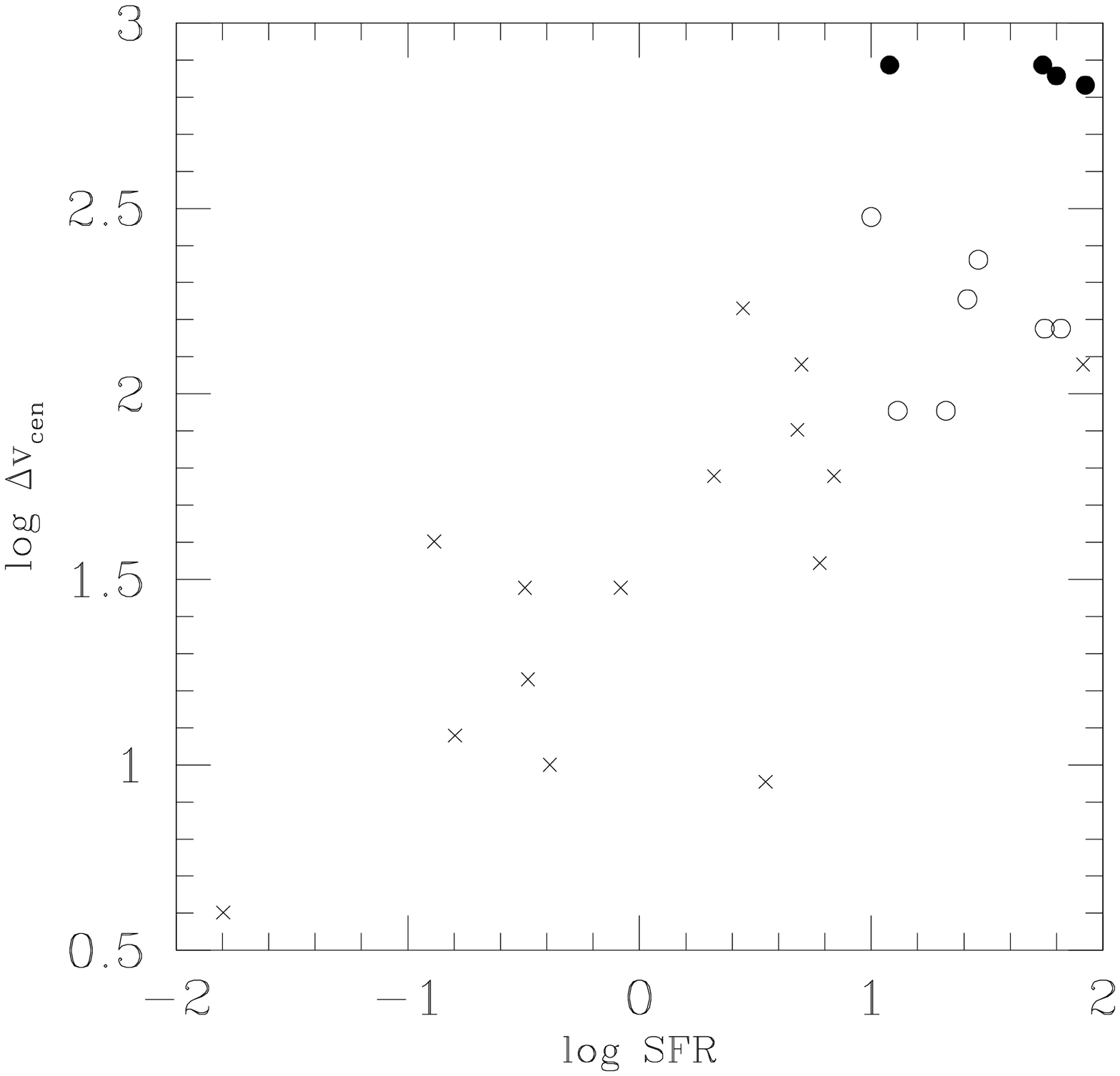} 
\caption{Plots of the outflow velocity 
(km s$^{-1}$) defined by the line centroid as a function of the galaxy stellar mass (M$_{\odot}$) and the star formation rate (M$_{\odot}$ yr$^{-1}$). The typical local stabursts are indicated by crosses and the Lyman Break Analogs with and without Dominant Central Objects are indicated by solid and hollow dots respectively. The exceptionally high outflow speeds associated with the Dominant Central Objects are clear. See Tables 1 and 2.}
\end{figure*}

These velocities are much higher than in the local starburst sample and in the LBAs
without a DCO. This is shown in Figure 7 where we plot the outflow speeds in the
ionized gas in local starburst and LBA samples {\it vs.}  the star formation rate and
galaxy mass (see Tables 1 and 2). For galaxies with FUSE data we use the
CIII$\lambda$977.0 and/or NII$\lambda$1084.0 and/or CII$\lambda$1036.3 lines.

The outflow velocities in the DCOs are also
significantly larger than those typically seen in high-z galaxies. Steidel et al. (2010) find
line centroids that are blueshifed on-average by 164 km s$^{-1}$ and maximum outflow
speeds that are typically $\sim$ 800 km s$^{-1}$ for galaxies with star formation rates of
$\sim 10^1$ to $10^2 M_{\odot}$ year$^{-1}$.

Outflows from intensely star-forming galaxies are believed to be produced as gas clouds
are accelerated by the ram pressure of a hot and fast wind driven by the collective
thermal/kinetic energy supplied by supernovae and massive stellar winds (e.g. Heckman
et al. 2000; Veilleux et al. 2005; Marcolini et al. 2005; Strickland \& Heckman 2010),
and/or by radiation pressure acting on dust (Murray et al. 2005; 2010). Can the unusually
high velocities seen in the DCOs be explained in this way? We consider a simple
idealized model of a cloud with a column density $N$ accelerated outward from an initial
radius $r_0$ by the ram pressure of a wind that carries momentum at a rate $\dot{p}$
into a solid angle $\Omega$. Then the terminal velocity of this cloud will be:\\

2) $v_{term} = 1800 \dot{p}_{35}^{1/2} (\Omega/ 4\pi)^{-1/2} r_{0,100}^{-1/2}
N_{21}^{-1/2}$ km s$^{-1}$\\

Here the momentum flux is in units of $10^{35}$ dynes, the initial radius is in units of 100
pc and the cloud column density is in units of $10^{21}$ cm$^{-2}$.  The DCO star formation
rates imply momentum fluxes of this order (Leitherer \& Heckman 1995), while the HST images
yield typical DCO radii of 100 pc (O9).
As discussed below in section 4.6, a column density of
$10^{21}$ cm$^{-2}$ is a reasonable estimate for the ionized gas in these objects.

The above neglects gravitational forces (cf. Wang 1995; Martin 2005; Murray et al. 2010;
Strickland \& Heckman 2010). This is justified here because the typical sizes and radii of
the DCOs imply virial velocities of only about 200 km s$^{-1}$ (much smaller than the
maximal outflow speeds we observe). Note that Murray et al. (2010) conclude that
outflows driven out of massive star clusters by radiation pressure will only have outflow
speeds of-order the local escape velocity. This is not the case in the DCOs.

These considerations imply that a galactic wind driven by an extreme starburst can in
principle accelerate clouds to the high velocities that we see in the DCOs. The
significantly higher velocities in these galaxies compared to typical starbursts can be
naturally explained by the combination of a large number of massive stars confined within
an unusually small radius. This leads to a very large wind ram pressure and/or radiation pressure
at the launch
point of the clouds.

\subsection{Feedback from an Active Galactic Nucleus?}

An alternative form of extreme feedback could be provided by an AGN. As discussed in
Overzier et al. (2009) the optical spectra of the LBAs with DCOs may imply
a composite of both an intense starburst and an obscured (Type 2) AGN. 
The COS data show that the far-UV radiation in these objects is dominated by light
produced by hot massive stars (Figures 2 and 3), which does not rule out the presence
of a Type 2 AGN.

Recent observations with XMM-Newton (Jia et al. 2010) show that LBAs with such apparently composite optical spectra are also over-luminous in the 2-10 keV X-ray band by factors of roughly three to thirty  compared to pure starbursts of the same far-IR luminosity.
They conclude that it is likely that at least some of these objects do harbor an AGN, but that the
overall bolometric luminosity of the galaxy is primarily due to an intense starburst. 
Could a Type 2 AGN be responsible for the unusually high outflow speeds and/or the 
apparently low covering factor of neutral gas due to a highly ionized state of the interstellar medium?

Very high outflow speeds (many thousands of km s$^{-1}$) are commonly observed in {\it unobscured} (Type 1)
AGN (e.g. Rupke et al. 2005; Krug et al. 2010). However these same authors show that outflows similar
to those we see in the DCOs (maximium outflow speeds of at least 1500 km s$^{-1}$) are extremely rare
in composite obscured-AGN/starburst systems and not seen at all in "pure" obscured AGN. They conclude
that the AGN can drive very high-speed outflows, but these flows are confined to a small region very near the
black hole (and are hence undetectable in Type 2 AGN). They also conclude that the AGN plays
little or no role in driving the large-scale outflows seen in the composite systems. It therefore seems
unlikely that the high velocity outflows in the DCOs are driven by an obscured AGN.

Tremonti et al. (2007) have detected outflows at velocities ranging from 500 to 2000 km s$^{-1}$ in massive
post-starburst galaxies at z $\sim$ 0.6. They speculate that such high velocities may require an AGN-driven
outflow. It is also possible that the these objects are more massive versions of the DCOs in this paper.
This could be confirmed by HST imaging of their galaxies.

\subsection{Clues from Lyman Alpha}

The properties of the Ly$\alpha$ profile are highly sensitive to the kinematics and
distribution of the outflowing gas and dust in star-forming galaxies (e.g. Steidel et al.
2010;  Hansen \& Oh 2006; Verhamme et al. 2006, 2008; Kornei et al. 2010).  In Figure 8 we plot the
Ly$\alpha$ profiles for the eight LBAs with COS data, and highlight one very
suggestive trend.

In the five cases in which the inferred relative escape fraction is small, there is very little
(if any) Ly$\alpha$ emission blueward of the systemic velocity. Instead, the profiles
show Ly$\alpha$ emission redward of the systemic velocity and either pure absorption or
a mix of absorption and weak emission to the blue. Such profiles are typical of high-z star
forming galaxies (e.g. Shapley et al.  2003; Steidel et al. 2010).  In contrast, for the three
objects in which we infer a high relative escape fraction (LBA0213+12, 0808+39, and
0921+45) a significant fraction of the Ly$\alpha$ emission is blueshifted with respect to
the systemic velocity. To quantify this, we have measured the net equivalent widths
(emission - absorption) blueward and redward of the systemic velocity in these eight
LBAs.  We define net emission (absorption) to have a positive (negative) equivalent
width. We then take the ratio of the red and blue equivalent widths ($R_{eqw}$).

\begin{figure*}
\includegraphics[trim = 0mm 0mm 0mm 0mm, clip,scale=.70,angle=-0]{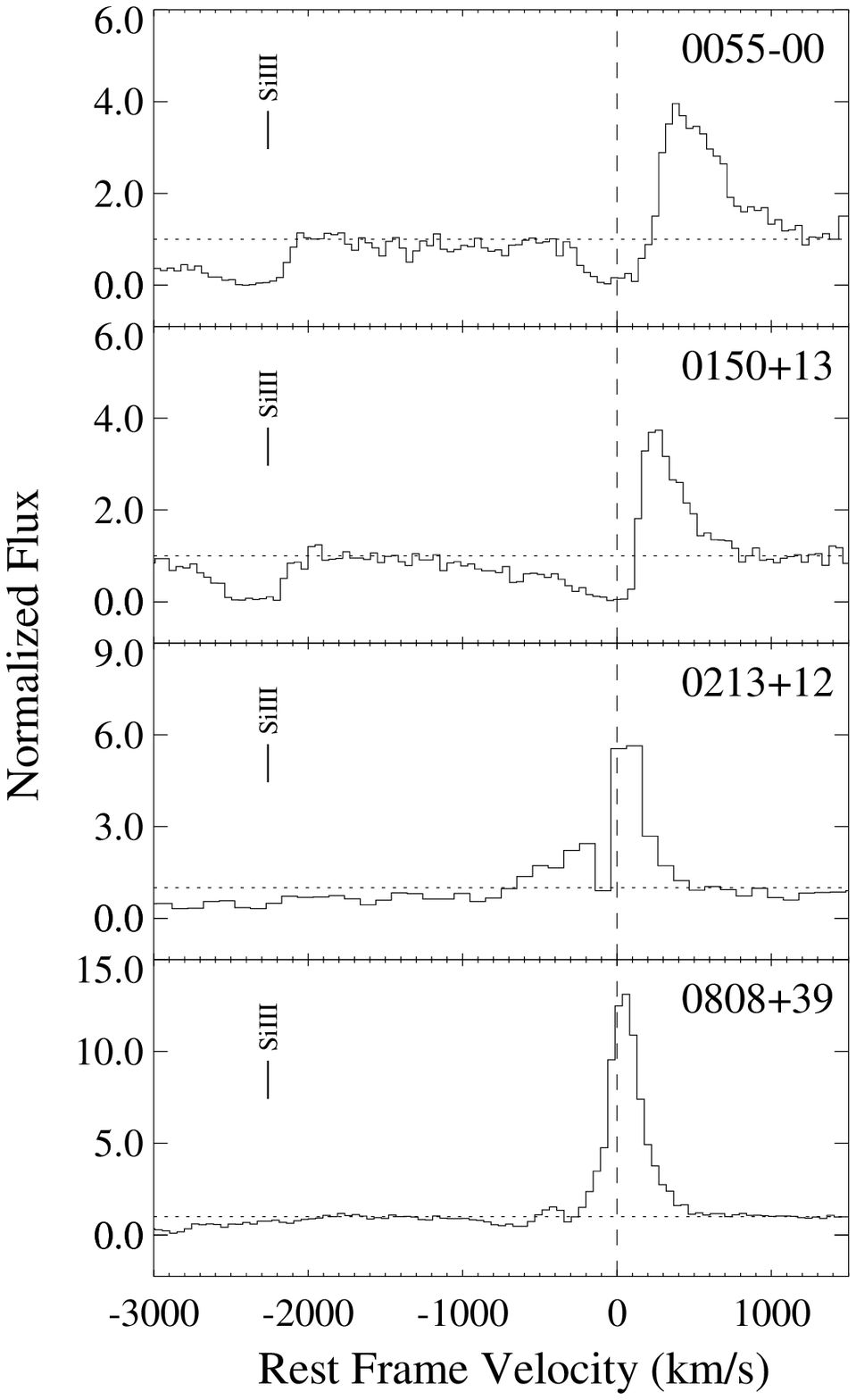} 
\includegraphics[trim = 0mm 0mm 0mm 0mm, clip,scale=.70,angle=-0]{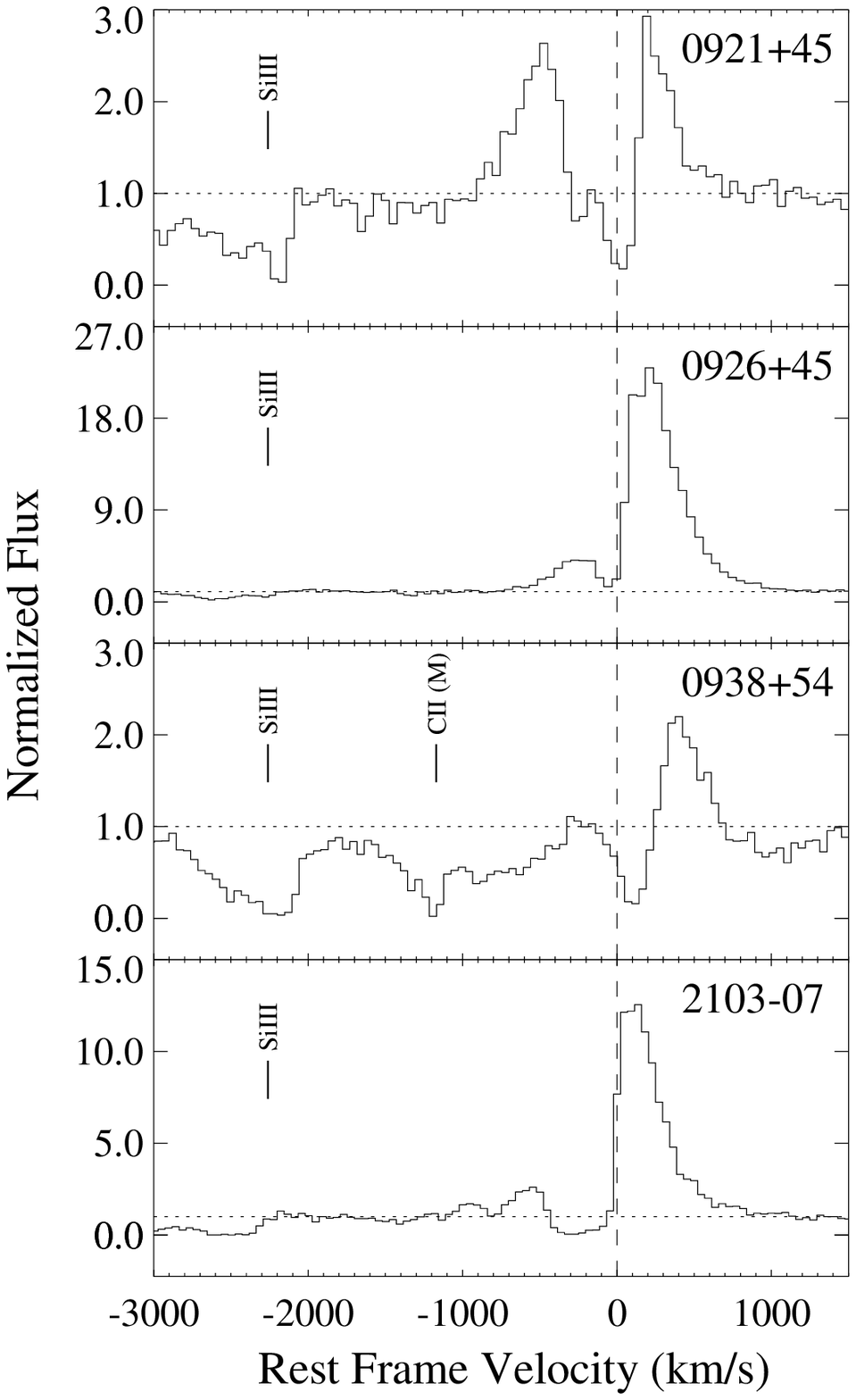} 
\caption{Plots of the Ly$\alpha$ feature. The location of the SiIII$\lambda$1206.5 feature is noted, as are the locations
of foreground Milky Way features. In three cases (0213+12, 0808+39, and 0921+45) there is a significant amount of blueshifted Ly$\alpha$ emission. These are the three objects that are candidates for a significant escape fraction of ionizing radiation.}
\end{figure*}

We list the results in Table 2. For the five galaxies with low inferred escape fractions
$R_{eqw}$ is either negative (redshifted emission and blueshifted absorption) or small
and positive (strong redshifted emission and weak blueshifted net emission):  LBA0055-
00 ($R_{eqw} =$ -0.21), 0150+13 (-0.78), 0926+45 (0.14), 0938+54 (-0.17), and 2103-07
(0.04). For the three galaxies with high inferred escape fractions $R_{eqw}$ is of-order one
and positive (similar amounts of red- and blue-shifted emission): 0213+12 ($R_{eqw} =$
0.36), 0808+39 (0.27), and 0921+45 (0.75).

Profiles with significant blueshifted emission are uncommon but not unknown in star-
forming galaxies at $z \sim$ 2 - 3. Verhamme et al. (2008) discuss two such cases and
interpret them as arising due to radiative transfer effects in a static medium. This model
can not be correct for the three DCOs  in Figure 8 because the metal lines demonstrate
that the gas is rapidly outflowing (Figures 4 and 6).  Erb et al. (2010) discuss the case of a high-z galaxy
with both an outflow detected in the metal lines and a significant amount of blueshifted
Ly$\alpha$ emission. They argue that in this case the optical depth to Ly$\alpha$ photons
is unusually low in the foreground outflowing material and that this is naturally related to
the low covering factor of neutral gas they infer in this object (precisely the situation we
see in our three galaxies).

\subsection{Where is the Dust?}

It is important to note that the observed UV spectral energy distributions (SEDs) of the
DCOs are significantly redder than the intrinsic SED of a starburst (Overzier et al.
2011, and see Table 2). For a normal starburst or SMC dust attenuation law (Calzetti 2001; Leitherer et
al, 2002) these SEDs imply roughly two magnitudes of dust extinction along the line-of-
sight to the DCO in the far-UV. This then raises a very interesting question. How can
there be this much dust when the observations of interstellar absorption-lines from the
neutral gas imply that such gas only covers a fraction of the far-UV source? The answer
must be that this dust is associated with {\it ionized} gas that covers most or all of the
DCO.

This picture requires that the cores of the strongest absorption-lines due to the ionized gas
should be nearly black (unlike the case of lines from the neutral phase). Examination of
the profiles in Figure 6 shows that this is the case in LBA0808+39 (a relative residual
intensity in the SiIII$\lambda$1206.5 line of 10\%) , LBA0921+45 (7\%), and LBA2103-
07 ($<$3\%).  In the case of LBA0213+12, the relative residual intensity in this line is
high ($\sim$40\%). However, the redshift of this galaxy is large enough to shift the
CIII$\lambda$977.0 line into the COS G130L spectrum. This should be the strongest
interstellar line due to ionized gas, and as such provides the most sensitive probe. Although the data are
noisy, the relative residual intensity in this line is small (10\% or less). Thus, in this galaxy the
SiIII line 
is optically thin, while the CIII line is optically thick.

We therefore conclude that the ionized gas has a near-unit covering factor in the DCOs.
One plausible physical model would be that the neutral gas represents the denser cores of
clouds being accelerated by the wind, while the ionized gas represents a lower density
halo of gas and dust surrounding the cloud core and
which is being ablated by the wind and photoionized by the intense radiation field (e.g.
Marcolini et al. 2005).

Can the DCO photoionize a column of gas with the required amount of dust? Elementary
considerations of photoionization equilibrium imply that a source of radiation can
photoionize a Stromgren slab with a column density given by:\\

3) $N_{Strom} = 1.2 \times 10^{23}~U$ cm$^{-2}$\\

Where $U$ is the dimensionless ionization parameter defined as the ratio of the density
of ionizing photons to electrons in the slab. In order for a significant fraction of ionizing photons to
escape, the actual gas column must be less than this value (to provide matter-bounded
conditions). The optical depth due to dust in the far-UV (1500 \AA) is related to the gas
column density by:\\

4) $N_{gas} = 3(5) \times 10^{20} \tau_{FUV} Z_{gas}^{-1}$ cm$^{-2}$\\

where the coefficients refer to an SMC (starburst) dust attenuation law (Leitherer et al.
2002) and we assume that the dust-to-gas ratio is proportional to the metallicity. 
The LBAs with DCOs all have gas-phase Oxygen abubdances close to solar (O09) and the
line-of-sight reddening implies $\tau_{FUV} \sim$ 2.
Thus, to
explain the line-of-sight reddening towards the DCOs in a matter-bounded situation we
need $U > 7 \times 10^{-3}$.

Alternatively, the optical depth at the Lyman edge due to neutral hydrogen in (mostly)
ionized gas with a column density $N_{HII}$ is given by :\\

5) $\tau_{Ly}  = 8.6 \times 10^{-24} N_{HII} U^{-1}$\\

For $N_{HII} = 10^{21}$ cm$^{-2}$, we require $U > 8 \times 10^{-3}$ for $\tau$$_{Ly}
< 1$.

Is this plausible? Heckman, Armus, \& Miley (1990) and Lehnert \& Heckman (1996)
showed that the observed radial gas density profiles seen in the optical emission-line
nebulae in starburst winds are consistent with theoretical expectations for photoionized
clouds exposed to  the ram-pressure of the hot wind fluid (Chevalier \& Clegg 1985). It is
straightforward to show that in this case the ionization parameter in the clouds is given
by:\\

6) $U = 9 \times 10^{-3}  Q_{55}/\dot{p}_{35}$\\

where $\dot{p}_{35}$ is the wind momentum flux in units of 10$^{35}$ dynes and
$Q_{55}$ is the starburst ionizing photon production rate (in units of 10$^{55}$ sec$^{-
1}$).

The ionizing photons are produced by the most massive stars with lifetimes $<$ 4 Myr,
while the momentum in the starburst wind derives primarily from supernovae (whose
progenitors have lifetimes of $<$ 40 Myr). In equilibrium (e.g.  a constant rate of star
formation for longer than 40 Myr), $Q_{55}/\dot{p}_{35} \sim$ 0.5 and  so $U \sim 5
\times 10^{-3}$. More generally, this ratio will be time dependent. For example, values
of $U  > 10^{-2}$ are associated with constant star formation for a duration less than 20
Myr and for an age $<$ 4 Myr in an instantaneous burst. These ages are consistent with
current constraints on the DCOs (O09 and see Figure 3)\footnote{We assume a standard
Chabrier (2003) initial mass function and use the Starburst99 models (Leitherer et al.
1999)}.

We therefore conclude that it is feasible for the DCOs to photoionize a matter-bounded
column of gas that is sufficient to contain enough dust to produce the observed reddening
and the inferred attenuation of the far-UV continuum and yet have a photoelectric
opacity at the Lyman edge that is less than one.

\section{Summary and Implications}

One of the major goals in cosmology is to understand the process by which the universe
was reionized. The most plausible source for the ionizing photons is an early generation
of hot massive stars in forming galaxies. The primary uncertainty at present is the
unknown fraction of such ionizing photons that are able to escape from the galaxies and
reach the neutral hydrogen in the inter-galactic medium. The column densities of gas
associated with such intense star formation correspond to extremely high optical depths
in the Lyman continuum. It seems nearly certain that some form of feedback from the
massive stars is required to produce the pathways through which the ionizing photons can
escape.

Our understanding of how this might occur would be greatly improved if we could find
local examples of galaxies from which significant ionizing radiation is escaping, and
could then investigate in some detail the physical processes that enable this to occur. This
will be most germane to the early universe if these local galaxies are good analogs to
intensely star-forming galaxies at high-redshift.

Accordingly, in this paper we used the Hubble Space Telescope Cosmic Origins
Spectrograph (COS) and the Far-Ultraviolet Spectroscopic Explorer (FUSE) to
investigate the far-UV properties of a sample of eleven Lyman Break Analogs (LBAs).
These rare local objects are very similar to high-z Lyman Break galaxies in all their
properties studied to date (Heckman et al 2005; Hoopes et al 2007; Basu-Zych et al.2007,
2009; Goncalves et al. 2010; Overzier et al. 2009 - O09; Overzier et al. 2008, 2010, 2011).
We have compared these rare objects to a sample of fifteen typical UV-bright local starbursts.

Following Heckman et al. 2001 (H01) and Grimes et al. 2009 (G09) we have used the
relative residual intensity in the cores of the strongest interstellar absorption-lines arising
in the neutral interstellar medium (the CII$\lambda$1036.5 and CII$\lambda$1334.5
lines) as a probe of the effects of the photoelectric opacity of the ISM on the Lyman
continuum. This approach is made possible by the high spectral resolution ($R \sim
10^4$) and good signal-to-noise in our data.

We find that there is a significant residual intensity ($\sim$ 30 to 50\%) in the CII line
core in four out of eleven of the LBAs but in none of the fifteen typical local starbursts.
We show that in all four of these cases the residual intensity reflects an
incomplete covering of the far-UV source by optically thick clouds (a picket fence
model).
Independent evidence for a significant escape of ionizing radiation
is provided by the anomalously weak H$\alpha$ emission (relative to the sum of the far-UV and far-IR
emission) in three of these four galaxies.  
{\it It is therefore plausible that a significant fraction of the Lyman continuum is
escaping in three of the LBAs in our sample}.

Assuming that the residual intensity in these lines does indicate the escape of ionizing
radiation, then following Shapley et al. (2006) and G09, the relative escape fraction in
these objects (defined as the ratio of fractions of escaping ionizing to non-ionizing far-
UV photons) is $\sim$30 to 40\%. After accounting for the effect of dust attenuation, the
absolute escape fractions are $\sim$ 3 to 10\%. 

We have also examined the profiles of the Ly$\alpha$ line in the COS data of the eight LBAs
for additional clues. We find that there is strong emission both blueward and redward of
the systemic velocity in the three objects in which a high escape fraction is likely. This
strongly suggests an incomplete covering of the FUV source by neutral clouds (in
agreement with the properties of the metal lines). The other five cases show the profiles
typically seen in high-z galaxies: Ly$\alpha$ emission redward of the systemic velocity
and either pure absorption or a mix of absorption and weak emission blueward.

In all three cases with high inferred relative escape fractions the HST images reveal that
most of the far-UV emission originates in a single dominant central object (DCO). These
are highly compact (radii $\sim 10^2$ pc) and massive (one to a few $\times 10^9$
M$_{\odot}$) young starbursts (O09). The interstellar absorption lines in all four of the
LBAs with DCOs show extremely high outflow speeds: the line centroids are blueshifted
by about 700  km s$^{-1}$ and the gas reaches maximum outflow velocities of at least 1500  km
s$^{-1}$. These are significantly higher than the velocities seen in other low redshift
starbursts or in typical high redshift galaxies. We have shown that these high speeds can
be explained in principle by the acceleration of gas clouds in a starburst-driven wind. The
exceptionally high velocities would be due to a large amount of momentum deposition in
such an unusually small region, leading to an extremely high wind ram pressure and/or radiation pressure at
the launch point for the clouds.

The far-UV emission from the DCOs is significantly reddened and attenuated by dust.
We have argued that this dust must reside in the ionized gas and have shown that the
ionized gas fully covers the far-UV source (the DCO). This situation is made possible by
the intense ionizing radiation field in these galaxies (leading to significant column densities of
dusty ionized gas that is optically thin in the Lyman continuum).

Based on the above considerations, it is tempting to link the escape of ionizing radiation
to the extreme feedback associated with the DCOs. The combination of a high rate of
energy deposition (kinetic energy and ionizing radiation) within such a small volume may
be the key to excavating channels through the interstellar medium and also intensely
irradiating these channels so that they are fully ionized and translucent to ionizing
radiation.

It is interesting that only 19\% ($6/31$) of the sample of LBAs imaged by O09 contain a
DCO. Together with the present results this would suggest that large relative escape
fractions are present in only 13\% of LBAs. This is similar to the fractions seen in high-
redshift galaxies (Shapley et al. 2006; Iwata et al. 2009).  Could the presence or absence of  DCOs
in these galaxies explain these results as well? 

It is important to note that in the sample of LBAs studied by
Overzier et al. (2009), the DCOs were preferentially found in the most massive galaxies
(M$_* \sim 10^{10}$ to $10^{11}$ M$_{\odot}$).  Galaxies this massive are exceedingly rare at $z >$ 7,
when the universe was reionized, and so the DCO host galaxies will not 
resemble the population of low-mass systems believed to be responsible for producing
most of the ionizing radiation at these early epochs (e.g. Bouwens et al. 2009). 

We are
instead suggesting that objects similar to the DCOs in this paper may exist in these early galaxies
and facilitate the escape of ionizing radiation. A number of theoretical papers over the past decade 
have investigated the processes that regulate the escape of Lyman continuum radiation from galaxies 
(e.g. Dove et al. 2000; Clarke \& Oey 2002; Fujita et al. 2003;
Gnedin \& Kravtsov 2008; Wise \& Cen 2009; Razoumov \& Somer-Larson 2010; Yajima et al. 2010). These studies are at least
qualitatively consistent with what we have observed in the DCOs.

Finally, we note that O09 have shown that the masses and sizes of the DCOs are
consistent with the regions of excess mass (cusps) seen in the centers of typical
present-day M$_*$
elliptical galaxies (Hopkins et al. 2009). If so, then such galaxies may have all passed
through an early stage during which they were sources of a significant amount of ionizing
radiation that was able to reach the intergalactic medium.



\acknowledgments



Facilities: \facility{HST(COS)}, \facility{GALEX}, \facility{SDSS}, \facility{Spitzer}

\end{document}